# First-Principles Study of Magnetism, Electronic Structure, and Bonding in Nb-Mn-Ge Kagome Compounds

Wei-Shen Tee[2,1], Shiya Chen[3], Weiyi Xia[1,2], Peter Minch[1,2], Cai-Zhuang Wang[1,2] and Vladimir Antropov[1,2]

[1]*Ames National Laboratory, U.S. Department of Energy, Iowa State University, Ames, Iowa 50011, USA*

[2]*Department of Physics and Astronomy, Iowa State University, Ames, Iowa 50011, USA*

[3]*Department of Physics, Xiamen University, Xiamen 361005, China*

## Abstract

In this work, we systematically investigate the magnetic ground states, electronic structures, and bonding characteristics of the computationally predicted stable $NbMn_6Ge_6$, $NbMn_6Ge_5$, and $NbMn_6Ge_4$ using first-principles calculations. Our results show that structurally stable $NbMn_6Ge_6$ has a collinear antiferromagnetic configuration, while the metastable rhombohedral $NbMn_6Ge_5$ and $NbMn_6Ge_4$ favor ferromagnetic ground states. Magnetic moments on Mn atoms are nearly localized, suggesting the applicability of a generalized spin Hamiltonian. Magnetic anisotropy in AFM $NbMn_6Ge_6$ and FM $NbMn_6Ge_5$ has uniaxial behavior, while FM $NbMn_6Ge_4$ has in-plane anisotropy. Charge density difference and electron localization function analyses further show charge redistribution and bonding features within the Mn-Ge Kagome network and between adjacent structural layers. Electronic structures near the Fermi level show no features suitable for topological magnetism studies. Experimental synthesis, structural characterization, and magnetic measurements are required to verify our predictions.

## Introduction

The *Kagome* lattice, named after a traditional Japanese woven bamboo basket pattern (*kago* meaning "basket" and *me* meaning "eye"), consists of a two-dimensional network of corner-sharing triangles that naturally forms hexagonal motifs. This unique geometry gives rise to geometrical frustration in the electronic hopping between neighboring atomic sites, leading to a wide range of intriguing electronic and magnetic properties, including flat bands [1-3], Dirac cones [3], van Hove singularities [4-6], nontrivial topological states [7], unconventional magnetism [8,9], anomalous Hall effects [10-12], and superconductivity [13-15].

The electronic structure of an ideal Kagome lattice is characterized by two dispersive bands intersecting at Dirac points and one nearly flat band. The flat band originates from destructive interference of electron hopping on the network of corner-sharing triangles, resulting in highly localized electronic states. In real Kagome materials, however, these characteristic bands are not always preserved. Their positions relative to the Fermi level and their degree of isolation depend sensitively on the chemical bonding, orbital symmetry, crystal-field splitting, and hybridization with neighboring atomic orbitals. Consequently, only compounds satisfying appropriate structural and chemical conditions exhibit well-defined Kagome bands [2,16].

Among the diverse families of Kagome compounds, the $AT_6X_6$ family (where A is a cation, T is a transition metal on the Kagome sites, and X is an element in the p-block) has recently attracted considerable attention because of its highly tunable structural [17], magnetic, and electronic properties. These compounds are composed of stacked transition-metal Kagome layers separated by nonmagnetic spacer atoms. Varying the spacer species tunes the interlayer coupling, magnetic exchange interactions, and electronic structure, yielding a wide range of magnetic ground states and topological electronic properties. Representative examples include V-based [18-21], Cr-based [22-26], Mn-based [3,27-30], Fe-based [31-37], and Co-based [36-40] compounds.

A recent high-throughput DFT study [17] of the $AT_6X_6$ kagome family predicted $NbMn_6Ge_6$ to be thermodynamically stable, with energy above the convex hull $E_d < 5$ meV/atom. Using a similar computational approach [41], $NbMn_6Ge_5$ and $NbMn_6Ge_4$ were identified as low-energy metastable phases with energies of 13 and 26 meV/atom above the convex hull, respectively. In general, the energy above the convex hull is commonly used as an indicator of thermodynamic stability. Compounds lying within approximately 20-30 meV/atom above the hull are often considered promising metastable candidates. A statistical analysis [42] of 29902 unique experimentally reported phases in the Inorganic Crystal Structure Database (ICSD) found a median metastability of approximately 15 meV/atom, while the 90th percentile is approximately 67 meV/atom. Experimentally accessible metastable phases can occur substantially farther above the hull, as their synthesis is governed not only by thermodynamic stability but also by kinetic factors, including reaction pathways [43,44], activation barriers [45], and practical experimental constraints. Therefore, the very small $E_d$ of Nb-Mn-Ge Kagome compounds provide strong thermodynamic support for its experimental accessibility and motivate further investigation of its structural, magnetic, and electronic properties.

In this work, we performed first-principles calculations to investigate the magnetic ground states, electronic structures, and chemical bonding of the earlier predicted stable $NbMn_6Ge_6$, $NbMn_6Ge_5$, and $NbMn_6Ge_4$ structures. We evaluated their thermodynamic stability in previous computational studies [17,41], and in the present work we calculated phonon dispersion relations to assess their dynamical stability. We determined the magnetic ground states by comparing several magnetic configurations. We analyzed band structures and density of states to reveal orbital contributions near the Fermi level, and we used charge density difference and electron localization function calculations to elucidate bonding characteristics and charge transfer.

## Results and Discussion

### A. Computational Methods

The first-principles density functional theory (DFT) calculations were performed using the Vienna *ab initio* simulation package (VASP) [46-48] with projector-augmented-wave (PAW) pseudopotentials [49] and Perdew-Burke-Ernzerhof (PBE) exchange-correlation energy functional [50]. Converged results are obtained with a kinetic energy cutoff for the plane wave basis of 520 eV and a $\Gamma$-centered k-point grid with a spacing of 0.2 $\text{Å}^{-1}$. The equilibrium structure is optimized until all forces are below 0.001 eV/Å. During structural relaxation, the Brillouin-zone integrations were performed using the Methfessel-Paxton smearing method with a smearing width of 0.1 eV. For subsequent static total-energy calculations, the tetrahedron method [51] was employed. For electronic density of states calculations, a denser k-point grid with a k-spacing of 0.1 $\text{Å}^{-1}$ was used. The structural models of $NbMn_6Ge_6$, $NbMn_6Ge_5$, and $NbMn_6Ge_4$ were taken from our previous computational studies [17,32] and were used here for further investigation of their magnetic and electronic properties.

The harmonic phonon calculations are performed using the finite displacement method implemented in the Phonopy code [52,53], based on spin-polarized forces obtained from VASP. The unit cells, consisting of 26 atoms for $NbMn_6Ge_6$, 72 atoms for $NbMn_6Ge_5$, and 33 atoms for $NbMn_6Ge_4$, are employed to capture the interatomic interactions.

### B. Crystal Structures

The $NbMn_6Ge_6$ structure predicted in Ref. [17] adopts the hexagonal $P6/mmm$ space group (No. 191), as shown in **Fig. 1(a)**. The structure is composed of alternating $Mn_3Ge$ Kagome, $Ge_2$, and $NbGe_2$ layers stacked along the *c* axis. Within the $Mn_3Ge$ layer (see **Fig. 1(b)**), Mn atoms occupy the vertices of the Kagome lattice with an in-plane Mn–Mn distance of 2.54 Å, while Ge atoms are located at the centers of the Kagome hexagons. Notably, the Ge atoms are displaced from the Mn Kagome plane along the *c* direction. The $Ge_2$ layer (see **Fig. 1(c)**) consists of Ge atoms arranged in a honeycomb lattice, whereas the $NbGe_2$ layer (see **Fig. 1(d)**) adopts a similar honeycomb framework in which Nb atoms occupy the centers of the Ge hexagons. Along the c-axis, the Mn Kagome layers stack directly above one another without any relative lateral displacement, preserving the crystal's high hexagonal symmetry.

Using a similar approach to Ref.[17], we identified $NbMn_6Ge_5$ and $NbMn_6Ge_4$ as low-energy metastable phases lying 13 and 26 meV/atom above the convex hull, respectively (for more details see Ref. [41]). Both $NbMn_6Ge_5$ and $NbMn_6Ge_4$ structures adopt the rhombohedral $R\overline{3}m$ space group (No. 166), which is closely related to the Li–Fe–Ge structure type [31,32,54]. In $NbMn_6Ge_5$, compared with $NbMn_6Ge_6$, the overall layered framework is preserved; however, the removal of one Ge atom lowers the crystal symmetry and introduces distortions into the Mn Kagome network. The previously equivalent Kagome triangles become inequivalent, producing alternating small and large triangles and breaking the ideal sixfold symmetry of the Kagome lattice. In addition, the Mn Kagome layers are no longer vertically aligned but shift laterally after every four layers, giving rise to an AABBCC stacking sequence (see **Fig. 2(a)**). In $NbMn_6Ge_4$, the stacking sequence differs from that of $NbMn_6Ge_5$: the Mn layers undergo a lateral shift after every two layers, resulting in an ABC stacking pattern, as illustrated in **Fig. 2(b)**. None of the three predicted crystal structures has yet been experimentally verified.

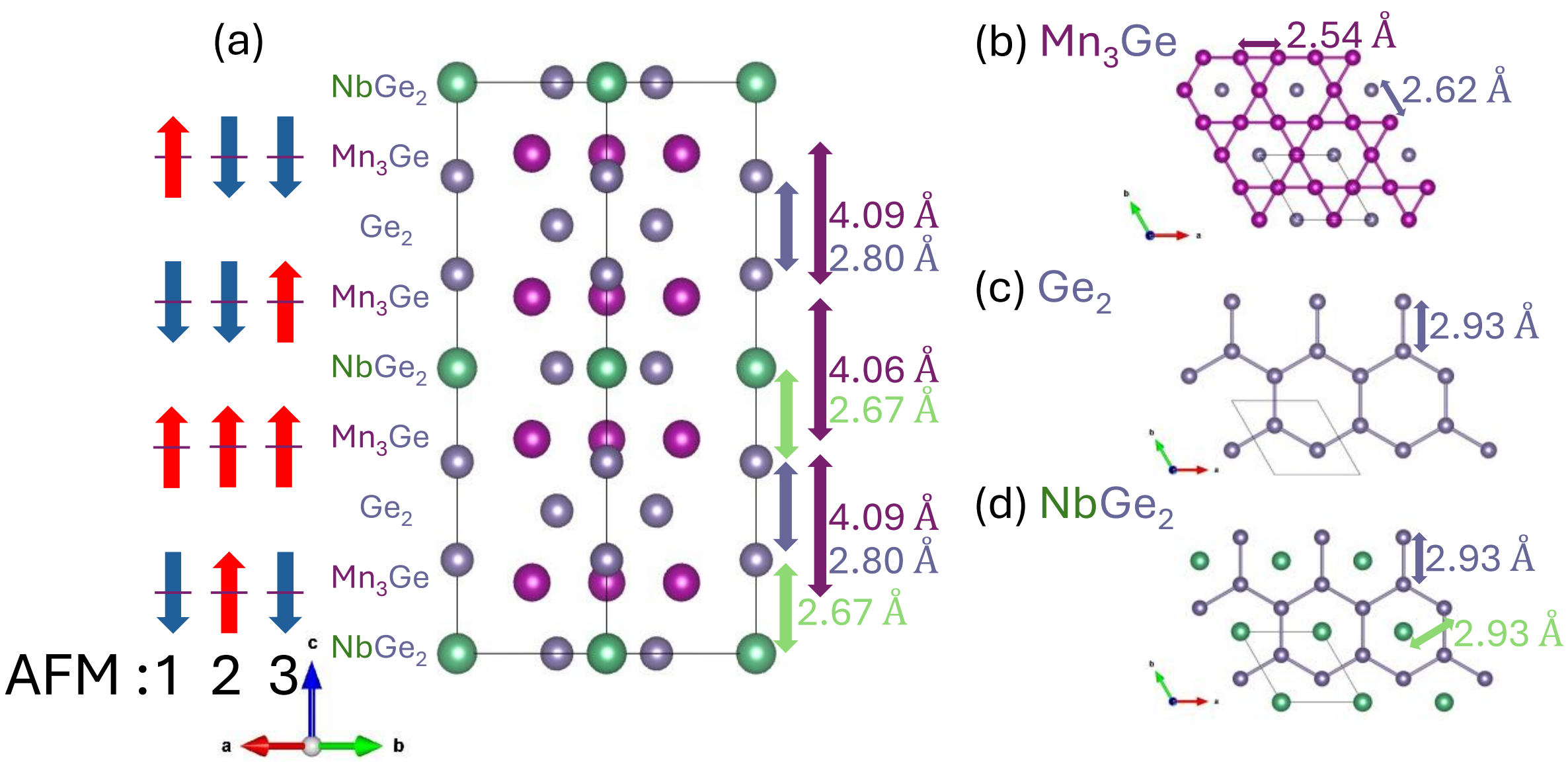

**Fig. 1** Crystal structures of **(a)** $NbMn_6Ge_6$ with the antiferromagnetic (AFM) spin arrangement adopted in the calculations and its constituent layers: **(b)** $Mn_3Ge$, **(c)** $Ge_2$, **(d)** $NbGe_2$ layer.

To determine the collinear magnetic ground state, each candidate's magnetic configuration was first fully relaxed, and its total energy was subsequently evaluated by a self-consistent calculation. The relative energies are summarized in **Table 1**. For $NbMn_6Ge_6$, the AFM2 configuration is found to be the magnetic ground state, while the FM, AFM1, and AFM3 states are higher in energy by 13.67, 54.92, and 37.18 meV/Mn, respectively. In contrast, $NbMn_6Ge_5$ and $NbMn_6Ge_4$ favor FM ground states, with the AFM configurations lying about 24-32 meV/Mn higher in energy. The calculated local Mn magnetic moments are generally around 2 $\mu_B$. The magnetic anisotropy energy shows that $NbMn_6Ge_6$ and $NbMn_6Ge_5$ has a weak uniaxial anisotropy, whereas $NbMn_6Ge_4$ favors an in-plane anisotropy.

| System | Electronic configuration | Energy (meV/Mn) | Atomic magnetic moment ($\mu_B$) | $E_x - E_z$ (meV/Mn) |
|---|---|---|---|---|
| $NbMn_6Ge_6$ | FM | 13.67 | 2.12, 2.24 | |
| | AFM1 | 54.92 | 1.94, 2.11 | |
| | AFM2 | 0 | 2.07, 2.20 | 0.07 |
| | AFM3 | 37.18 | 1.96, 2.11 | |
| $NbMn_6Ge_5$ | FM | 0 | 2.07, 2.21 | 0.03 |
| | AFM1 | 32.21 | 1.99, 2.22 | |
| | AFM2 | 29.19 | 1.92, 2.32 | |
| $NbMn_6Ge_4$ | FM | 0 | 2.15, 2.29 | -0.13 |
| | AFM1 | 24.14 | 2.03, 2.23 | |
| | AFM2 | 26.19 | 2.04, 2.23 | |

**Table 1.** Calculated energies, local Mn magnetic moments, and magnetic anisotropy energies ($E_x - E_z$) for different magnetic configurations of $NbMn_6Ge_6$, $NbMn_6Ge_5$, and $NbMn_6Ge_4$.

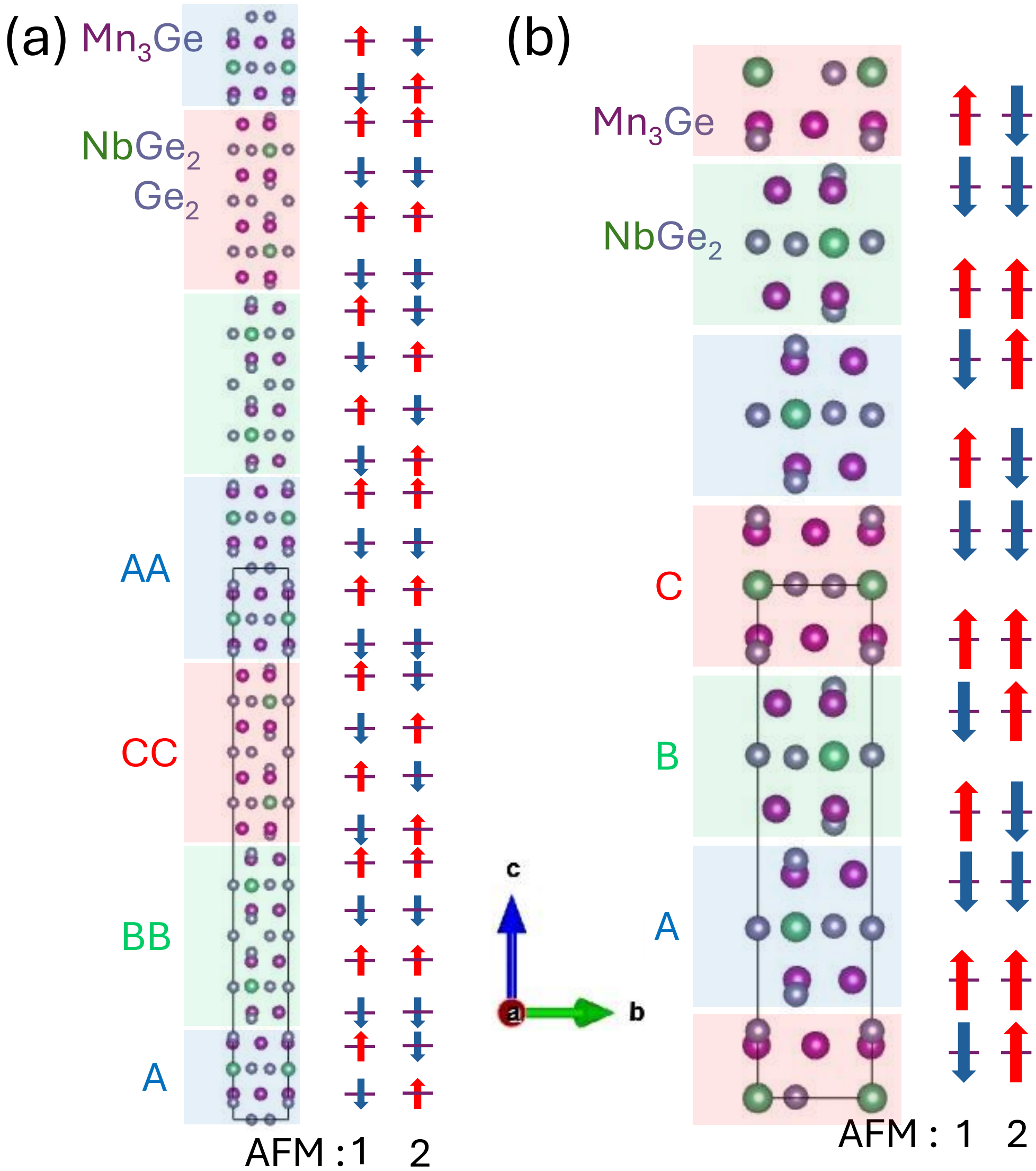

**Fig. 2** Crystal structures of **(a)** $NbMn_6Ge_5$ and **(b)** $NbMn_6Ge_4$ with the AFM spin arrangement adopted in the calculations. Colored boxes highlight Mn Kagome layers with identical in-plane positions.

### C. Phonon Dispersion and Dynamical Stability

We examine the dynamical stability of the optimized structures through phonon calculations. **Fig. 3** shows the calculated phonon dispersion relations along the high-symmetry paths of the Brillouin zone. No imaginary phonon frequencies were observed for $NbMn_6Ge_6$, $NbMn_6Ge_5$, and $NbMn_6Ge_4$, indicating these three structures are dynamically stable at T = 0 K. Phonon dispersion relations have been reported for several members of the $AT_6X_6$ kagome family in previous studies [17,21,55], providing useful comparisons for the lattice-dynamical behavior of the present compounds.

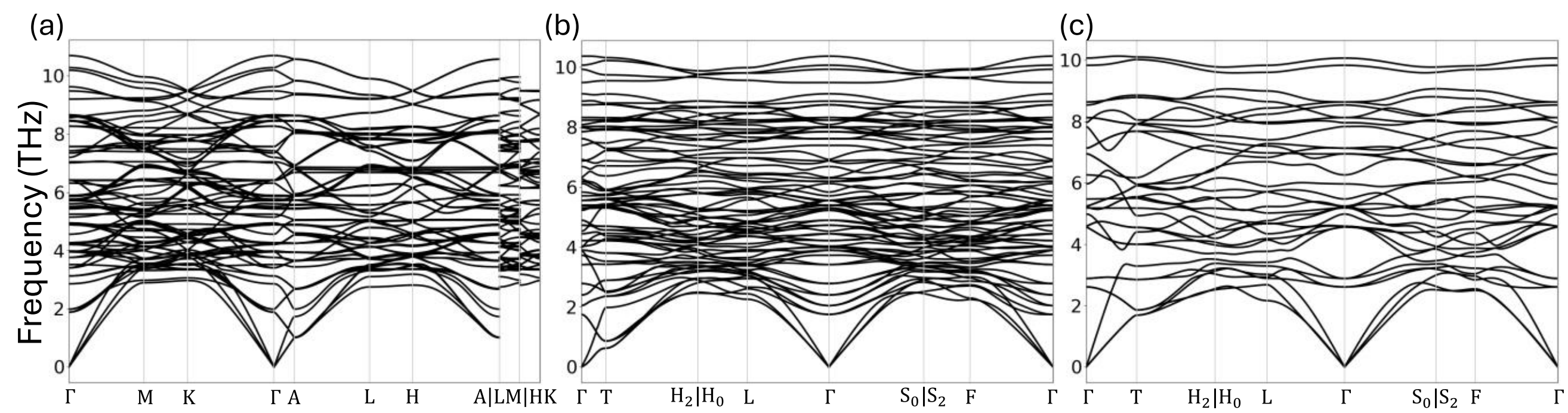

**Fig. 3** Phonon dispersion relation of **(a)** $NbMn_6Ge_6$, **(b)** $NbMn_6Ge_5$, and **(c)** $NbMn_6Ge_4$.

## D. Electronic structure analysis

The band structures of three Nb-Mn-Ge compounds are shown on **Fig. 4**. The calculated band structures reveal that the states near the Fermi level are dominated by Mn 3d orbitals. Based on their symmetry relative to the Kagome plane, the d orbitals are grouped into three sets: the mainly in-plane $d_{xy}$ and $d_{x^2-y^2}$ orbitals, the $d_{xz}$ and $d_{yz}$ orbitals with mixed-in-plane and out-of-plane character, and the $d_{z^2}$ orbital oriented primarily along the vertical direction. A clear spin dependence is observed. In AFM2 $NbMn_6Ge_6$, the spin-up and spin-down bands is broadly similar dispersions, whereas in FM $NbMn_6Ge_5$ and $NbMn_6Ge_4$, the spin-up bands show more pronounced flattening while the corresponding spin-down bands remain relatively dispersive. In addition, $NbMn_6Ge_5$ exhibits another nearly flat feature along the $\Gamma - \mathrm{T}$ direction with predominantly $d_{z^2}$ orbital character. The in-plane $d_{xy}$ and $d_{x^2-y^2}$ bands are observed at approximately 0.77 eV above the Fermi level in the $NbMn_6Ge_6$, $NbMn_6Ge_5$, and $NbMn_6Ge_4$ structures. Interestingly, these bands become nearly flat along the $\Gamma - \mathrm{T}$ direction in $NbMn_6Ge_5$ and $NbMn_6Ge_4$, while remaining more dispersive in $NbMn_6Ge_6$.

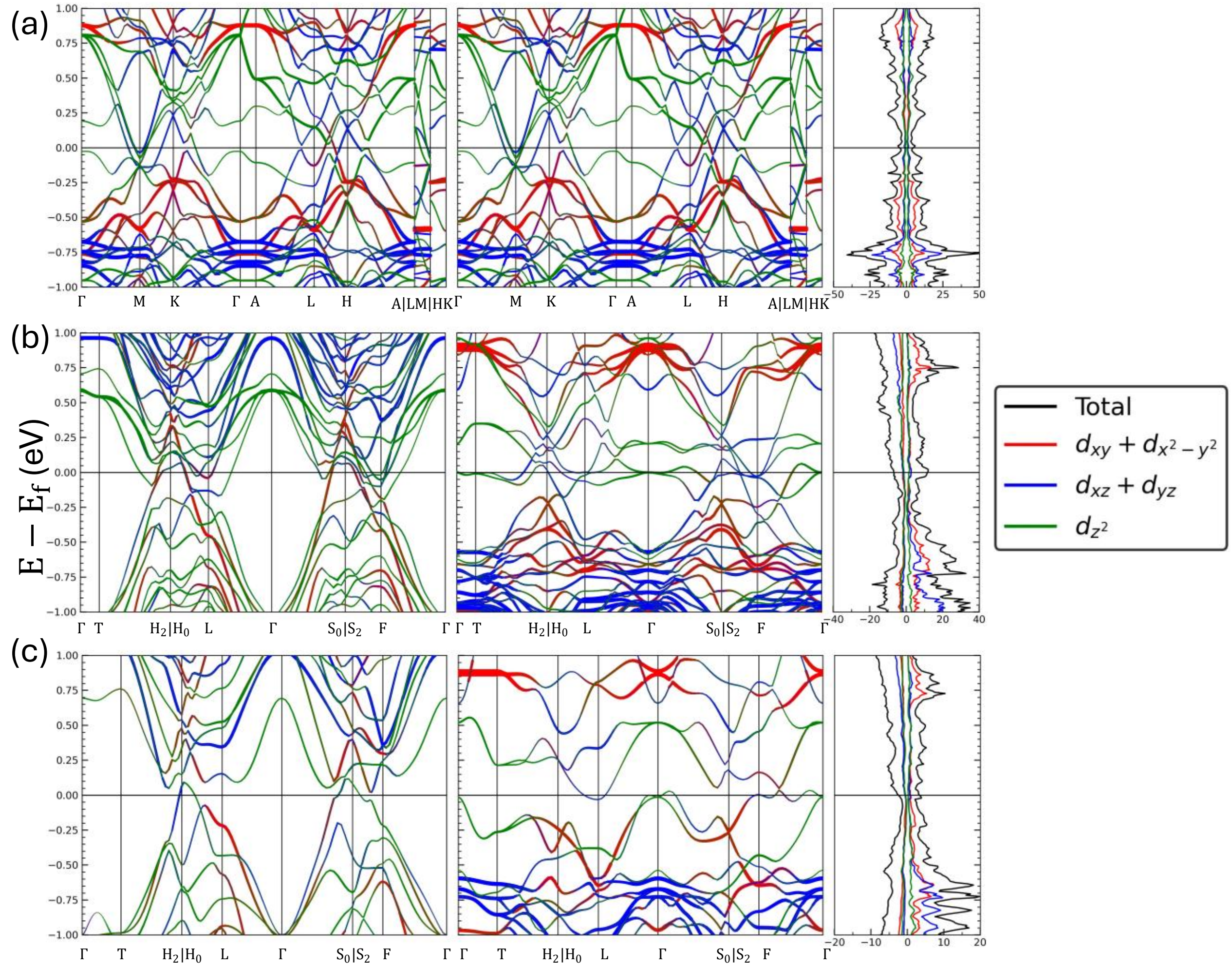

**Fig. 4.** Spin-polarized orbital-resolved electronic band structures and densities of states (DOS) of **(a)** $NbMn_6Ge_6$, **(b)** $NbMn_6Ge_5$, and **(c)** $NbMn_6Ge_4$. $NbMn_6Ge_6$ is shown in its AFM2 ground state, whereas $NbMn_6Ge_5$ and $NbMn_6Ge_4$ are shown in their ferromagnetic (FM) ground states. For each compound, the minority-spin band structure is shown in the left panel, the majority-spin band structure in the middle panel, and the total DOS in the right panel. The electronic bands are projected onto the Mn 3d orbitals, with $d_{xy} + d_{x^2-y^2}$ shown in red, $d_{xz} + d_{yz}$ in blue, and $d_{z^2}$ in green. The Fermi level is set to 0 eV.

Nonrelativistic and relativistic bands are shown in **Fig. 5** for comparison. The effects of spin-orbit coupling (SOC) are more pronounced along the out-of-plane $\Gamma - \mathrm{A}$ direction, while the overall band dispersions remain largely unchanged. SOC lifts the band degeneracy at the $\Gamma$ point, resulting in a clear splitting of the corresponding bands. The relatively small changes elsewhere in the Brillouin zone indicate that the main electronic structures are preserved upon inclusion of SOC. This suggests that these electronic features are primarily governed by lattice geometry and orbital hybridization, while SOC acts mainly to lift specific degeneracies and modifies the electronic structure near the $\Gamma$ point.

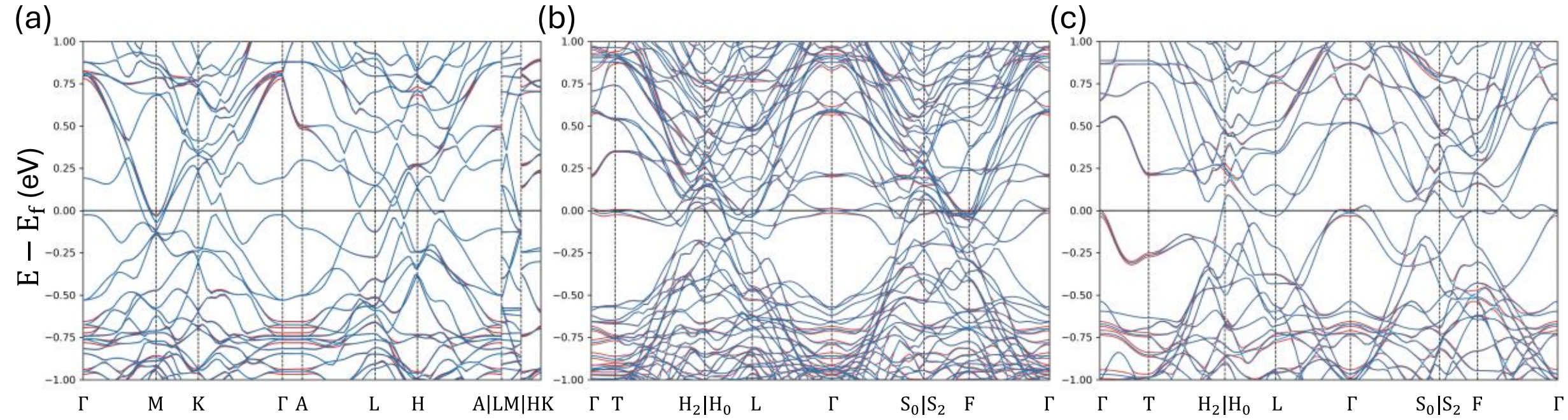

**Fig. 5.** Comparison of the electronic band structures calculated with and without spin-orbit coupling (SOC) for **(a)** $NbMn_6Ge_6$, **(b)** $NbMn_6Ge_5$, and **(c)** $NbMn_6Ge_4$. The blue and red bands represent the calculations without and with SOC, respectively. The Fermi level is set to 0 eV.

The charge density difference (CDD) and electron localization function (ELF) of $NbMn_6Ge_6$ are shown in **Fig. 6** and **Fig. 7**, respectively. The CDD reveals significant charge accumulation within the $Mn_3Ge$ kagome layer, particularly in the triangular regions formed by three neighboring Mn atoms. However, the corresponding ELF distribution does not exhibit strong electron localization in these regions, indicating that the accumulated charge is largely delocalized within the kagome network. These delocalized electronic states are consistent with the metallic character of the compound.

In contrast, a highly localized electron density is observed in the Ge-Ge dumbbell (see **Fig. 7(a)** red box). This feature can be interpreted as a Ge–Ge dumbbell containing lone-pair-like electron states. A similar Ge-Ge dumbbell has also been reported in $MgCo_6Ge_6$ [40], $CaFe_6Ge_6$ and $CaCo_6Ge_6$ [36], and $LuCr_6Ge_6$ [25].

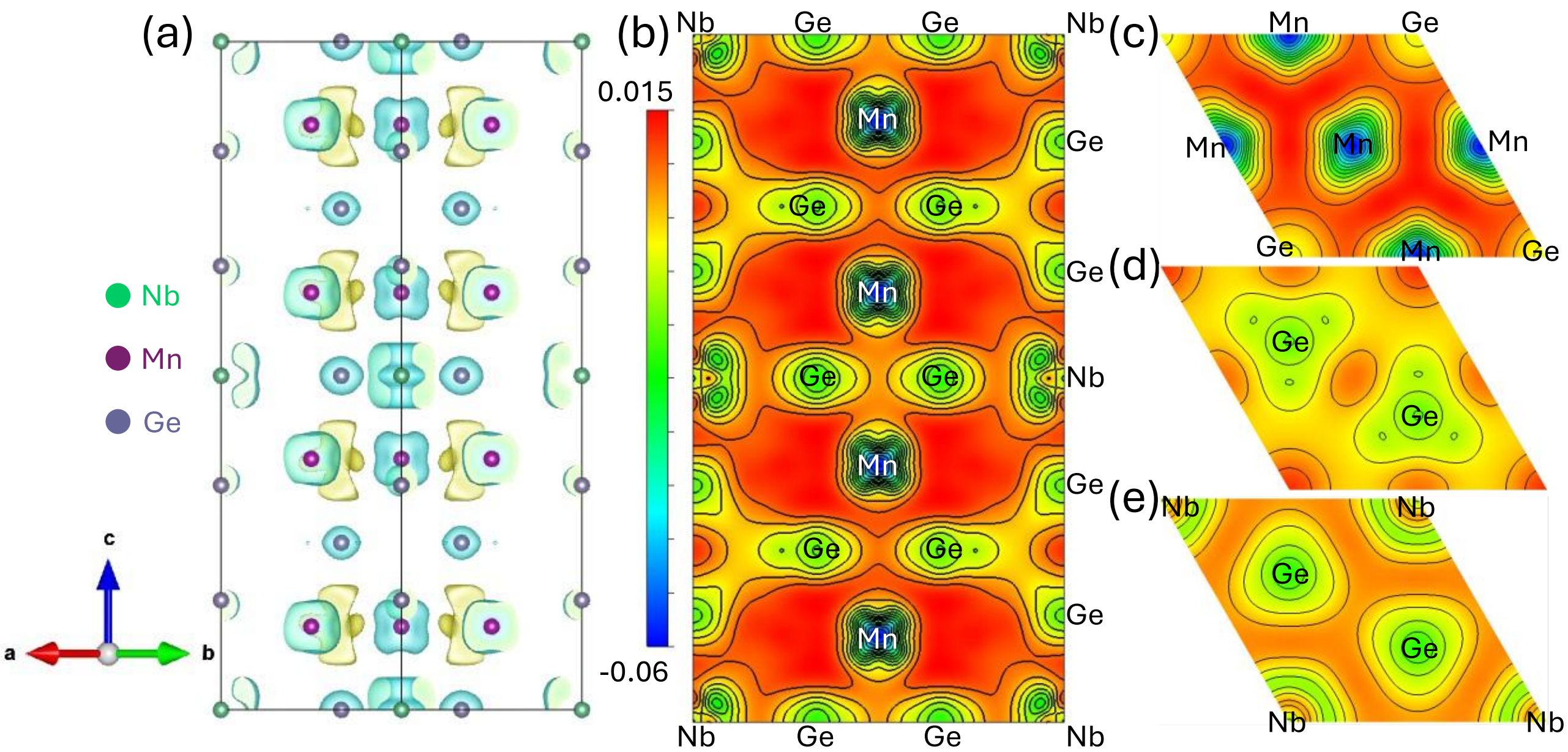

**Fig. 6.** Charge density difference (CDD) of $NbMn_6Ge_6$. **(a)** Three-dimensional isosurface view of the CDD with an isovalue of $\pm 0.01\ e$, where yellow and blue regions represent charge accumulation and charge depletion, respectively. **(b)** CDD viewed along the (110) plane. **(c)** CDD in the $Mn_3Ge$ Kagome plane. **(d)** CDD in the $Ge_2$ honeycomb plane. **(e)** CDD in the $NbGe_2$ plane.

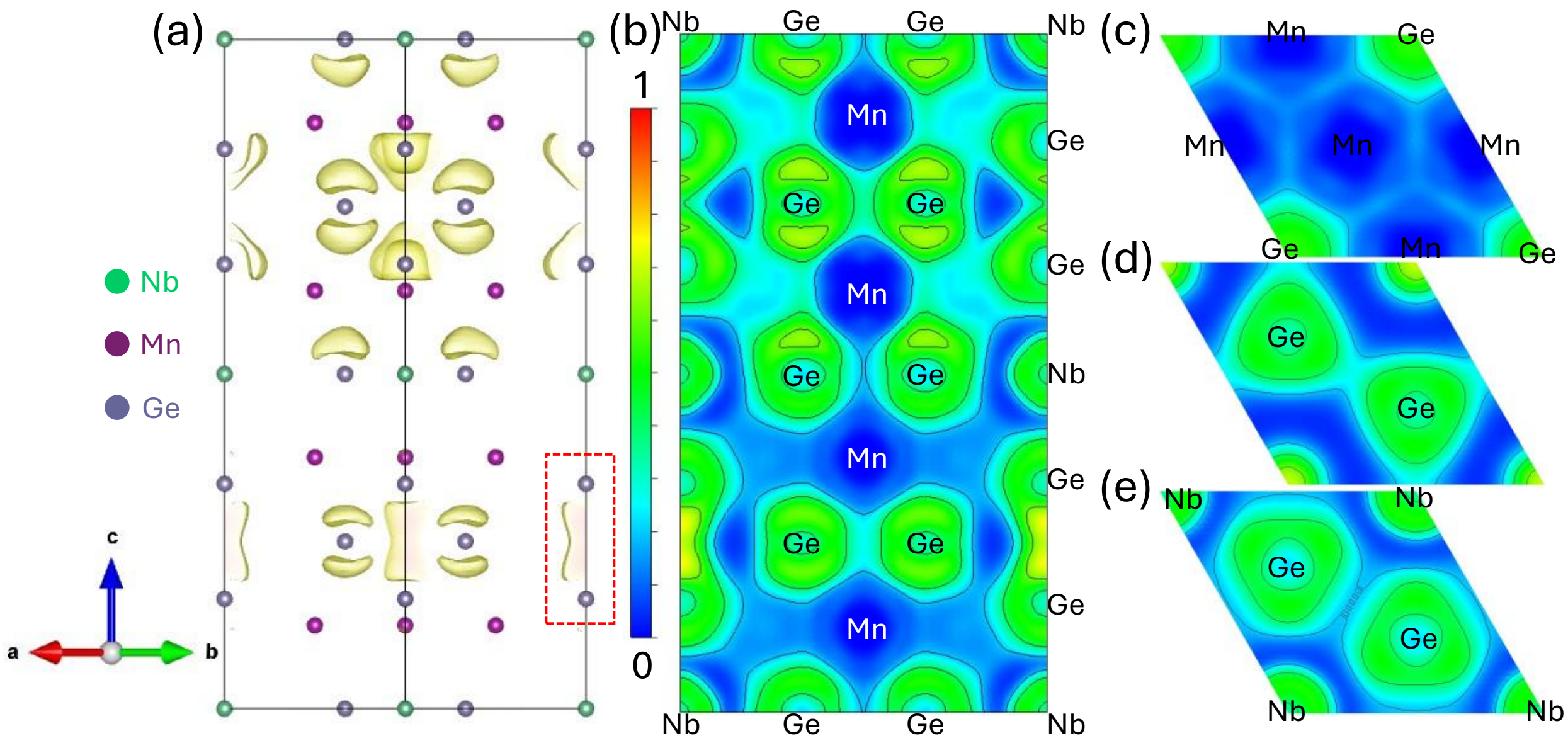

**Fig. 7.** Electron localization function (ELF) of $NbMn_6Ge_6$. **(a)** Three-dimensional ELF isosurface of the crystal structure with an isovalue of 0.55, where the red box indicates the Ge-Ge dumbbell. **(b)** ELF viewed along the (110) plane. **(c)** ELF in the $Mn_3Ge$ Kagome plane. **(d)** ELF in the $Ge_2$ honeycomb plane. **(e)** ELF in the $NbGe_2$ plane. Higher ELF values indicate stronger electron localization, whereas lower ELF values correspond to more delocalized electronic states.

## Conclusions

In summary, we have studied the crystal and magnetic structure stability, electronic structures, charge density difference, and electron localization function of $NbMn_6Ge_6$, $NbMn_6Ge_5$, and $NbMn_6Ge_4$ using first-principles calculations. Predicted $NbMn_6Ge_4$ and $NbMn_6Ge_5$ structures adopt the rhombohedral $R\overline{3}m$ space group. We also find that $NbMn_6Ge_6$ has an AFM ground state, while removing Ge atoms stabilizes an FM ground state in $NbMn_6Ge_5$ and $NbMn_6Ge_4$. Magnetism in these systems appears to be local, so we suggest that a generalized spin Hamiltonian applies. The energy difference between the FM and the closest AFM state increases after Ge atoms are removed from $NbMn_6Ge_6$, so one can expect a corresponding increase of the Curie temperature. The magnetic anisotropy in FM $NbMn_6Ge_4$ appears to be in-plane. Experimental synthesis, structural characterization, and magnetic measurements are required to verify these predicted structures.

## Acknowledgements

Work at Ames National Laboratory was supported by the U.S. Department of Energy (DOE), Office of Science, Basic Energy Sciences, Materials Sciences and Engineering Division, including a grant of computer time at the National Energy Research Scientific Computing Center (NERSC), Berkeley, CA. Ames National Laboratory is operated for the U.S. DOE by Iowa State University under Contract No. DE-AC02-07CH11358. Shiya Chen was supported by the National Natural Science Foundation of China (Grant No. T2422016) and the Natural Science Foundation of Xiamen (Grant No. 3502Z202371007).

## Conflicts of Interest

There are no conflicts to declare.